\documentclass{article}

\usepackage{xcolor}
\usepackage{setspace}
\usepackage{float}
\usepackage[]{changebar}
\usepackage{mathtools}
\usepackage{amsmath,amssymb,amsfonts}
\usepackage{amstext}
\usepackage{graphicx}
\usepackage{stfloats}
\usepackage{authblk}
\usepackage[]{url}

\usepackage{xcolor}
\usepackage{array}
\usepackage[]{changebar}
\usepackage[final]{changes}
\colorlet{Changes@Color}{red}
\definechangesauthor[color=red]{R1}

\newcommand{\rev}[1]{\added[id=R1]{#1}}

\title{Thermal vulnerability detection in integrated electronic and photonic circuits using IR thermography}

\author[1*]{Bilal Hussain}
\author[2*]{Bushra Jalil}
\author[2]{Maria Antonietta Pascali}
\author[1]{Muhammad Imran}
\author[1]{Giovanni Serafino}
\author[2]{Davide Moroni}
\author[1]{Paolo Ghelfi}

\affil[1]{Institute of Technologies for Communication, Information and Perception, Scuola Superiore Sant'Anna, Pisa, Italy}
\affil[2]{Istituto di Scienza e Tecnologie dell’Informazione “Alessandro Faedo” CNR, 56124 Pisa, Italy}

\affil[*]{Corresponding author: bilal.hussain@santannapisa.it, bushra.jalil@isti.cnr.it}

\begin{document}

\maketitle

\begin{abstract}
Failure prediction of any electrical/optical component is crucial for estimating its operating life. Using high temperature operating life (HTOL) tests, it is possible to model the failure mechanisms for integrated circuits. Conventional HTOL standards are not suitable for operating life prediction of photonic components owing to their functional dependence on thermo-optic effect. This work presents an IR-assisted thermal vulnerability detection technique suitable for photonic as well as electronic components. By accurately mapping the thermal profile of an integrated circuit under a stress condition, it is possible to precisely locate the heat center for predicting the long-term operational failures within the device under test. For the first time, the reliability testing is extended to a fully functional microwave photonic system using conventional IR thermography. By applying image fusion using affine transformation on multimodal acquisition, it was demonstrated that by comparing the IR profile and GDSII layout, it is possible to accurately locate the heat centers along with spatial information on the type of component. Multiple IR profiles of optical as well as electrical components/circuits were acquired and mapped onto the layout files. In order to ascertain the degree of effectiveness of the proposed technique, IR profiles of CMOS RF and digital circuits were also analyzed. The presented technique offers a reliable automated identification of heat spots within a circuit/system. 
\end{abstract}

%\setboolean{displaycopyright}{true}

\section{Introduction}

Recent advances in photonic integrated circuit (PIC) technologies have ushered a new era in IC technology \cite{jalali2006silicon}, \cite{hochberg2013silicon}. There has been a huge interest in silicon photonics due to its compatibility with existing complementary metal-oxide semiconductor (CMOS) technology paving the way for monolithic integration of electronic and photonic components \cite{jalali2006silicon}. The systems/subsystems comprising of integrated electronic and photonic components with relatively smaller footprint, are cost-effective and energy-efficient.  However design and development of reliable photonic-electronic systems is a challenging task. The PICs are functionally different from conventional CMOS technologies in several aspects. For example, in contrast to conventional ICs, PICs contain very few metal layers and have relatively limited number of components consuming DC currents \cite{patel2015silicon}, \cite{maram2019recent}. The operating currents of optical components are on the order of few mA whereas for conventional CMOS ICs the current consumption can reach up to few Amperes \cite{kim201316}, \cite{essing201427ghz}. Also, most of the signal elaboration in the optical domain such as modulation and filtering is achieved using thermo-optic effect inside the semi-conductor material. Therefore, precise temperature control and monitoring of photonic integrated circuits become paramount in a microwave photonic system design. The strong dependence of photonic components on ambient temperature render conventional failure modeling ineffective. 

For conventional CMOS technologies, the physics of failure (PoF) is dependent on the effective failure modeling of mechanisms that ultimately causes the breakdown of the component. Instead of operating the IC for an extended period of time, the operating life is estimated by subjecting the device under test (DUT) to added stress (thermal or electrical). Such reliability testing is commonly known as high temperature operating life (HTOL) test governed by JEDEC standards \cite{mcpherson2018brief,breitenstein2004thermal,breitenstein2011luminescence,breitenstein2013illuminated,breitenstein2004lock}. In HTOL tests, the devices are subjected to elevated temperatures (higher than the room temperature) in a controlled environment and the device performance is monitored \cite{mcpherson2018brief}. The exposure to high temperatures induces thermal stress and thereby helps in predicting the operating life of a device/component. For example, for electronic systems the stress is induced by applying a voltage or by increasing the ambient temperature in a test chamber. The device performance is monitored over a specific time interval and the failure conditions are extracted for the device under test (DUT) . These failure conditions are further used to model the failure mechanism as well as the PoF. Using HTOL tests, failure mechanisms such as electromigration (EM), time-dependent dielectric breakdown (TDDB), negative bias temperature instability (NBTI) and hot carrier injection (HCI), are modeled. The failure mechanism can degrade the device performance that may eventually lead to a complete breakdown/failure. 

Conventional HTOL testing is not suitable for PICs due to their strong dependence on temperature. If a PIC is subjected to an elevated operating temperature, the device will cease its function and thereby rendering conventional HTOL test incapable of providing sufficient information on the performance of the DUT. There exists a strong need \rev{for} developing modified HTOL testing suitable for photonic components. Therefore instead of exposing/subjecting PICs to conventional HTOL conditions, in-operation thermal profiling in conjunction with performance monitoring can be used to identify failure mechanisms and predict its operating life i.e., long term reliability.  A flow chart of IR-assisted operating life reliability test is presented in Fig. 1. \rev {There exist several techniques for obtaining thermal profile of a component/system. Conventionally, the most basic form of temperature monitoring is using a thermistor. For optical components, a micrometer sized thermistor is placed on a convenient location over a PIC. Using electrical connections, it is possible to monitor the localized temperature change with an accuracy of 0.01 K. This technique is not suitable for heat profiling of a complete PIC due to its limited spatial resolution. Recently, other non-contact heat profiling techniques such as thermal reflectance microscopy \cite{Liu} , two photon absorption \cite{Zhuo} and IR thermography \cite{Breitenstein2011ThermalFA} are presented for thermal profiling of circuits. Thermal reflectance monitors the temperature by using thermo-modulation of the refractive index of the material. A laser point source is used to monitor the changes in the reflectance of the material under observation. The temperature variation is computed by derivating the reflected spectra. This technique is more suitable for sub-micron laser cavities due to presence of large temperature gradients on the cavity facets. For very large surfaces such as silicon based optical circuits, thermal reflectance is not an efficient technique of heat profiling due to the fact that devices are on the order of millimeters and cannot be scanned with optimum speed. Another non-contact thermal profiling technique is based on two photon absorption in fluorescent rhodamine dye (R6G). The CMOS device is coated with a layer of R6G and near infrared range fluorescence is monitored using a microscope. The fluorescence is caused by two photon absorption of lower energy photons (Far IR range). This technique is recently used for monitoring CMOS devices such as memory chips. As the fluorescence is dependent on photon absorption thus requires a strong excitation due to low probability of simultaneous excitation by two photons. This technique is not suitable for monitoring silicon based optical devices  due to low metal density and low optical powers involved. Another thermal profiling technique, thanks to the advancement in far IR receivers, is using an IR camera. This non-contact method can provide thermal profile of a device with an accuracy of 1mK (with lock-in). The limited size of a pixel on the un-cooled detector limits the spatial and temperature resolution of the thermogram. As sub-micron resolution is not needed for silicon based optical devices, IR camera based thermal profiling can be employed for thermal monitoring of PICs due to speed, cost-effectiveness and simple handling. A comparison of various thermal profiling techniques is presented in Table \ref{tab:Techniques} }

Mere mapping of thermal profile is not sufficient for predicting the operating life of a PIC. The location of heat centers where thermal energy is accumulated needs to be correlated with the actual layout of the device. Thermal profile mapping onto the layout of PIC can be achieved using innovative computer vision techniques.

% \begin{table}[ht]
% \centering
% \caption{\rev{\bf A comparison of various thermal profiling techniques}}
% \begin{tabular}{m{6em}  m{4em} m{6em}  m{2em} m{1em}}
% \hline
% Thermal Profiling Technique & Spatial Resolution & Temperature & Non Contact &  Cost \\
% \hline
% Thermistor based Profiling & Limited & 0.01 K & No & + \\
% Infrared Camera & Limited (Pixel size)  & 4K without lock-in 0.001K with lock-in & Yes & ++ \\
% Thermal reflectance & Few nanometers & 0.5K (strongly dependent on material)& Yes & +++ \\
% Two photon absorption & >800 nm & 1K but requires strong excitation & Yes & +++\\
% \hline
% \end{tabular}
%   \label{tab:Techniques_old}
% \end{table}

\begin{table}[ht]
\centering
\caption{\rev{\bf A comparison of various thermal profiling techniques}}
\begin{tabular}{>{\raggedright\arraybackslash}m{5em}  >{\raggedright\arraybackslash}m{5em} >{\raggedright\arraybackslash}m{6em}  >{\raggedright\arraybackslash}m{2em} >{\raggedright\arraybackslash}m{2em}}
\hline
Thermal Profiling Technique & Spatial Resolution & Temperature & Non Contact &  Cost \\
\hline
Thermistor based Profiling & Limited & 0.01 K & No & + \\
Infrared Camera & Limited (Pixel size)  & 4K without lock-in 0.001K with lock-in & Yes & ++ \\
Thermal reflectance & Few nanometers & 0.5K (strongly dependent on material)& Yes & +++ \\
Two photon absorption & >800 nm & 1K but requires strong excitation & Yes & +++\\
\hline
\end{tabular}
  \label{tab:Techniques}
\end{table}

\subsection{Failure prediction using mid-IR imaging: State of the art}
Computer\rev{-aided} fault detection and monitoring is already implemented in the electronics industry for printed circuit board (PCB) technologies. An automatic PCB inspection method consists of fault detection and possible errors in bare or assembled PCBs e.g. break\rev{s} in circuits, missed components, wrong polarity etc.  Initially, these methods develop reference or calibrated data set and later the inspected board is compared with the reference. At first, before an image can be used as a source of information, it is of extreme importance to perform necessary pre-processing to obtain uniform and stable results. Pre-processing performs different techniques to enhance the features of the image however it does not increase the image information. Different steps involved in pre-processing steps include camera calibration, light balancing, deblurring followed by edge detection and filtering. 

\begin{figure}[htbp]
 \centering
  \fbox{\includegraphics[width=.4\textwidth]{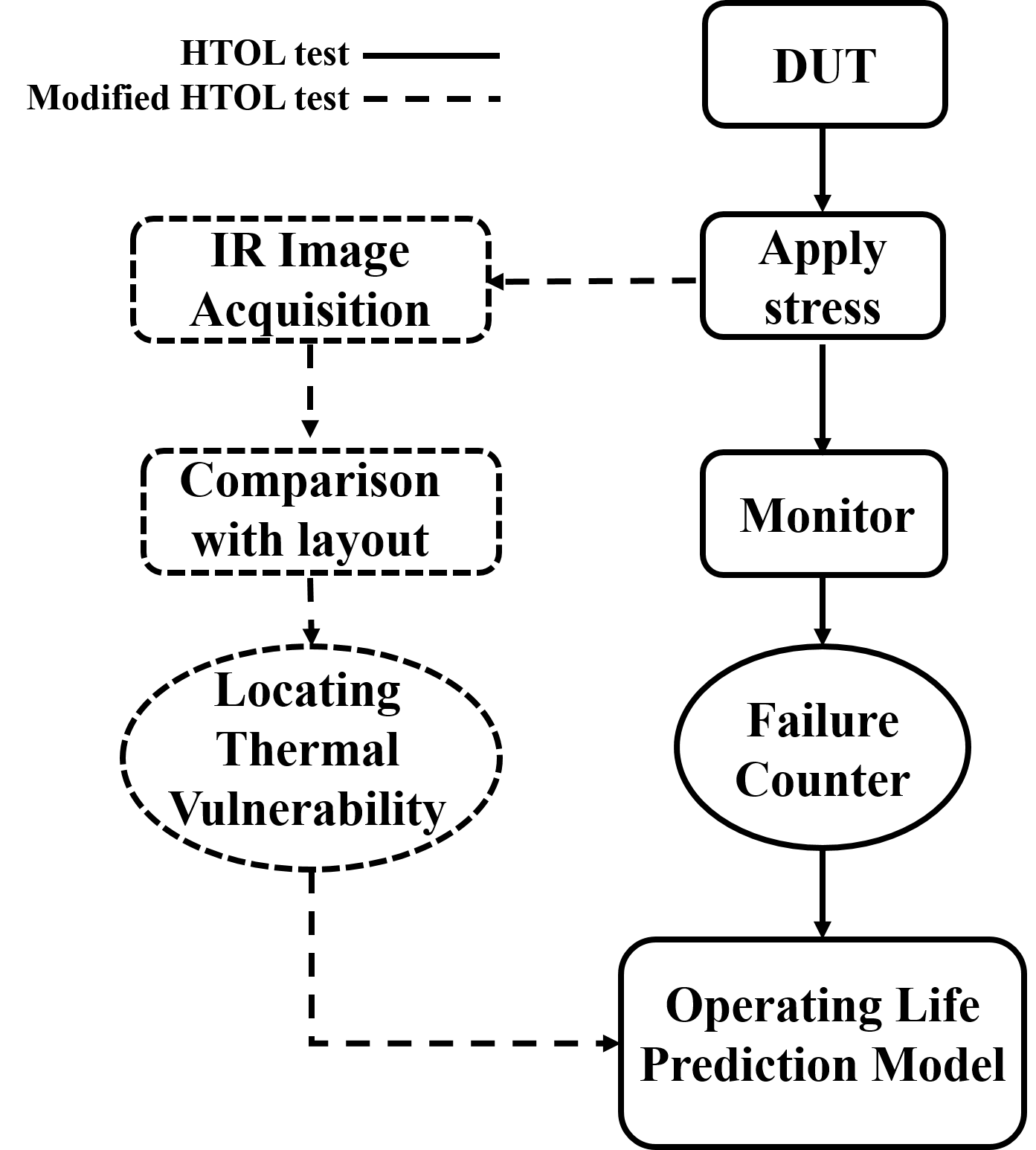}}
    \caption{IR-assisted HTOL reliability testing.}
  \label{fig:figure_1}
\end{figure}

After pre-processing, different approaches to identify potential faults are used. Most of these methods convert RGB to \rev{Grayscale} and perform thresholding or morphological operations. Some of these algorithms examine geometrical or shape features and perform module matching and detection of components in assembled PCB. Several other methods are based on different computer vision theories e.g. image segmentation or edge detection, Hough transform, watershed segmentation and wavelet\rev{-}based approaches \cite{Santoyo, Furat,kim,Wagh}. However, in most of the cases\rev{,} these methods have lower detection accuracy and ultimately suffer \rev{from} the problem of identification of fuzzy edges, corners, noise and precise segmentation. 

\begin{figure}[!b]
\centering
   \includegraphics[width=.85\textwidth]{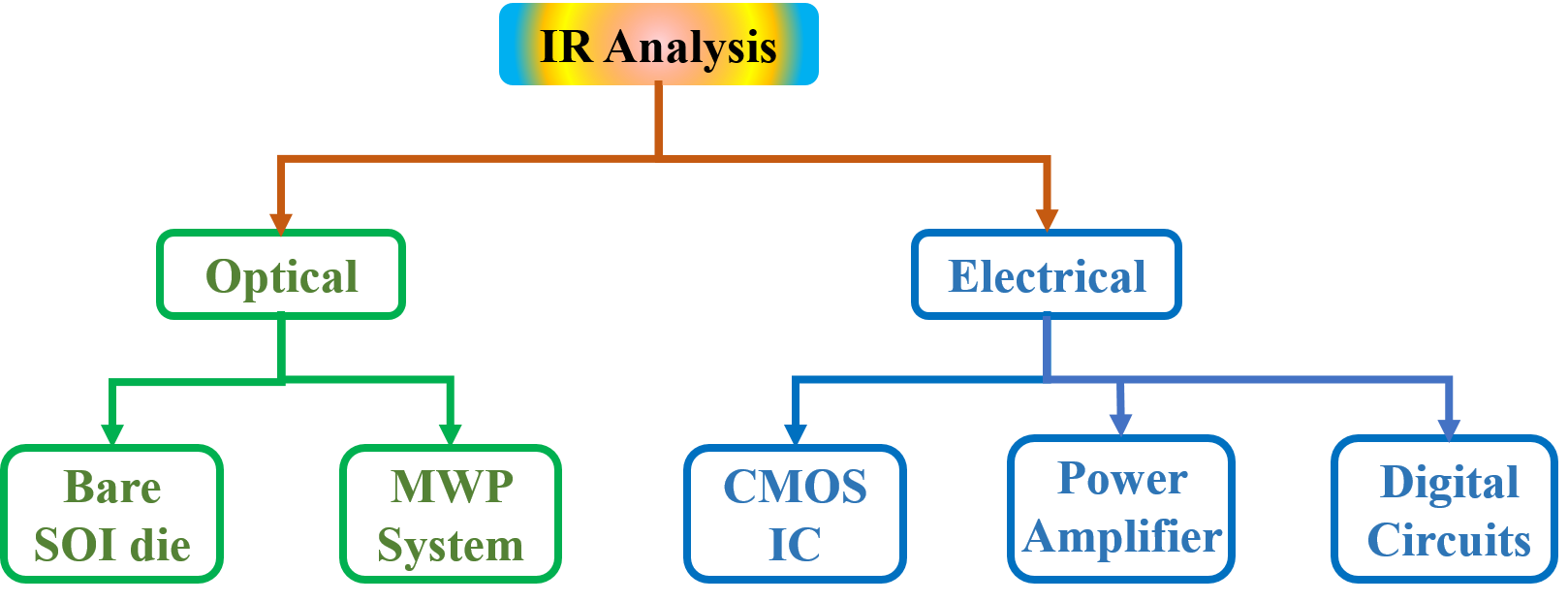}
    \caption{Types of circuits analyzed using proposed IR-assisted thermography technique.}
  \label{fig:figure_2}
\end{figure}

Thermal and infrared imaging of PCBs came up as an alternative solution to test and inspect electronic components. The thermal effects  are  one  of  the  main  limiting  factors  in  electronic equipment  operation. The IR profiling of electronic component\rev{s}  is  necessary  to detect  their possible  failures  and  to  enhance  their reliability. Several techniques based on these multimodal sensors \rev{have} been proposed to inspect and identify fatal errors from these boards. The main challenge of these approaches is to find an algorithm that can perfectly align the test and real PCB image. Ibrahim et al. presented a method that provides a well\rev{-}aligned defective image to enhance a PCB inspection algorithm with real images \cite{ibrahim}. Recently, Anitha et al. in their paper presented a detail\rev{ed} survey of different computer vision methods used to detect defects in different categories of PCBs such as, single layer, double layer and multilayer Bare PCB and Assembled PCB \cite{Anitha}.  Based on their survey, they highlighted that most of these methods emphasize on bare PCB images to identify common faults on single, double or multi-layer PCB images. The limitation of such techniques is that it cannot locate faults within the hidden layers of a PCB. For PCB as well as integrated \rev{technologies}, an automated technique is needed such that it can model the faults using the layout information of PCB/IC.

Inspired from automated PCB inspection, similar IR profiling techniques can be extended for photonic circuits. Instead of using visible range images/die photograph of PIC, the layout files (GDSII format \rev{ for integrated circuits and Gerber format for PCB technologies}) can be used as a reference image for accurate heat profiling across the PIC surface. An accurate heat profiling can improve \rev{the} reliability modelling of PICs by precisely locating heat centers along the circuit.
\begin{figure}[t]
\centering
  \includegraphics[width=.75\textwidth]{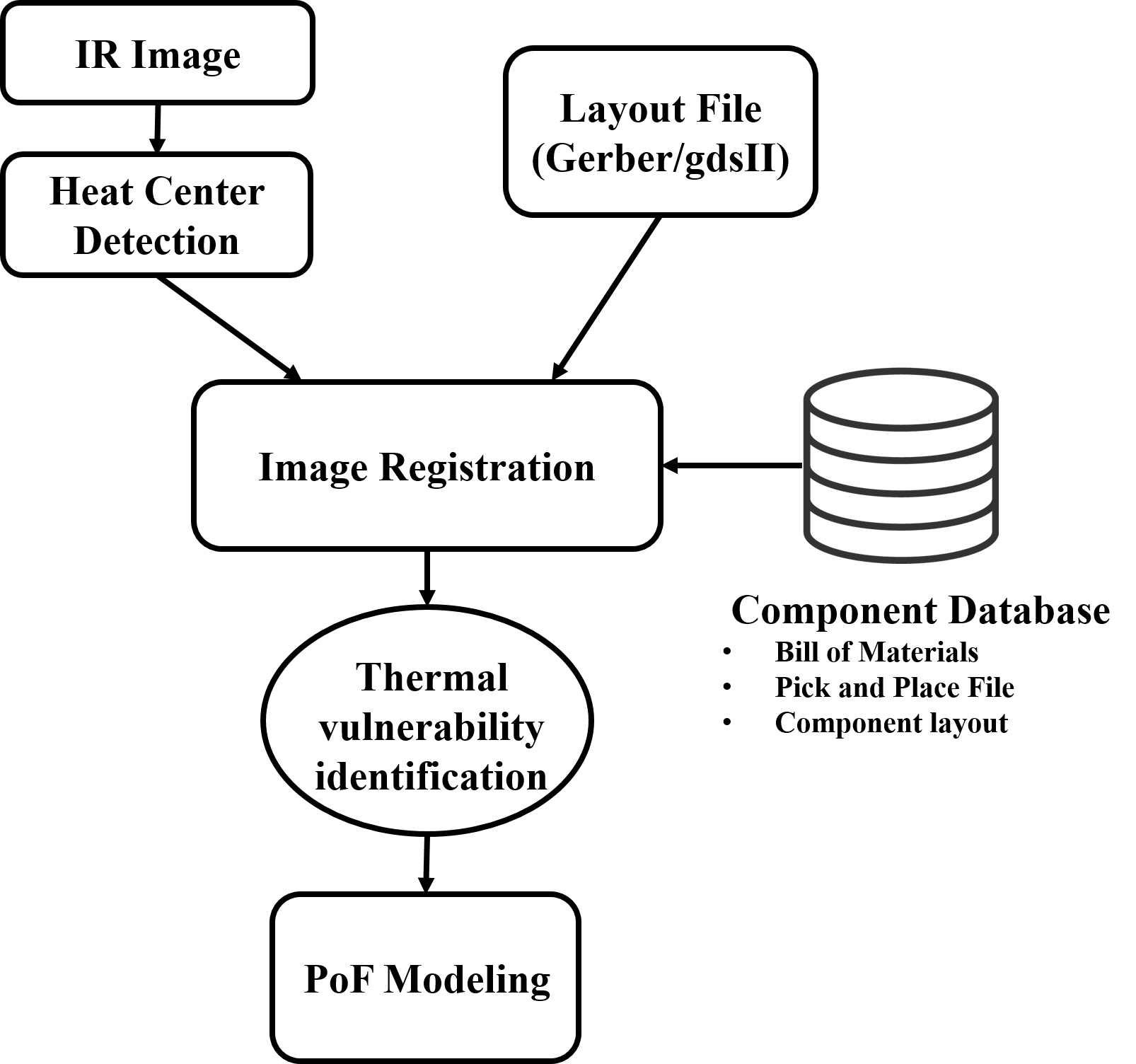}
    \caption{Illustration of proposed methodology.}
  \label{fig:figure3}
\end{figure}

In this work, we applied an automated in-operation thermal profiling technique suitable for future photonic integrated circuit technologies. The proposed method can be used to identify failure mechanisms and predict device operating life i.e., long term reliability without subjecting PICs to conventional HTOL conditions. The technique can be easily augmented with performance monitoring  for reliable device/system failure prediction. We have developed an IR profiling technique which is based on comparison of GDSII or gerber files instead of relying on visible range images. The algorithm processes the acquired IR profile of the DUT and then compares it with its layout. After verifying the proposed technique on complete PCB, IR profiles of various  electronic and photonic circuits were analyzed to further understand the strengths and limitations of the proposed technique as depicted in Fig. 2. At first, the thermal profile of a PIC is analyzed.  Furthermore, a functional microwave photonic system is analyzed using the proposed IR profiling technique. The IR thermography is also extended to CMOS RF and digital circuits. Various integrated as well as discreet PCB circuits are analyzed. For PCB technologies, we have used multilayer PCBs and have shown that the proposed technique is able to predict heat center(s) along with identification of active component on the PCB.  The presented IR profiling technique can be useful for performing HTOL tests as the IR thermography is non-invasive in nature.
\begin{figure}[!b]
\centering
 
  \includegraphics[width=.75\textwidth]{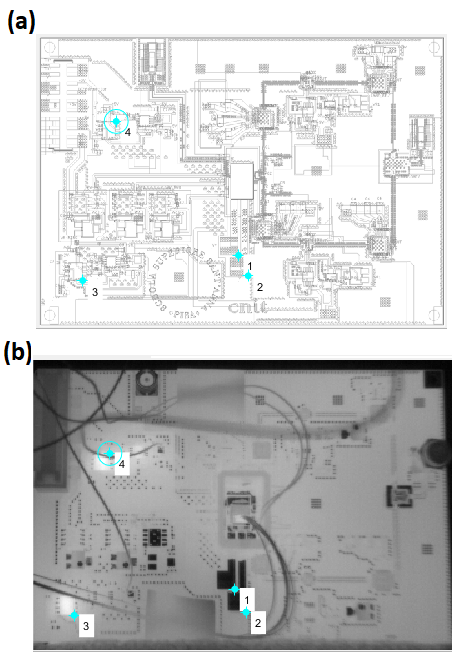}
    \caption{a) Gerber image of the top layer of PCB design, b) Thermal profile of circuit board top layer.}
  \label{fig:4point}
\end{figure}

 In the next section, we will explain the experimental setup and data acquisition followed by the applied thermal imaging techniques. The detailed explanation of image registration is provided in section 2. Section 3 explains the thermal analysis and findings  of different optical and electronic circuits. Section 4 concludes this work and identifies open issues for failure modeling in photonic components. 
 
\section{Materials and Methods}

\subsection{Data Acquisition}

Infrared data and images in the visible spectrum have been acquired under different conditions. The infrared camera used in the acquisition set-up is the FLIR A65sc \rev{(FLIR Systems Inc.),
%approximate cost 7.000\$ at the purchase time in 2017,
a radiometric camera using an uncooled VOX microbolometer detector  and equipped} with the standard lens  (25mm focal length, H-FOV x V-FOV = 25$^{\circ}$ x 20$^{\circ}$)\rev{; this camera has a thermal sensitivity of 0.05°C and a spatial resolution, i.e., instantaneous field of view (IFOV), of 0.68 mrad. The data was acquired at the frame rate about 5f/s with the infrared camera. The FLIR A65 camera produces IR images with 640 x 512 pixels. During the experiments in some cases the unwanted background was present and in that case we cropped those images for better visualisation and to focus only on the interest area. By cropping the images, the original thermal values were not changed, only the size was reduced}. In order to map the heat profile of 4-layer Rogers 4350b PCB (size 8$\times$ 11.5cm) containing electronic as well as photonic components, the FLIR camera is mounted at a distance of approx 9 cm while keeping the whole system in a thermally insulated container (to avoid external heat sources). 
\rev{With this camera-target distance, the H-FOV is of 0.04 m and the V-FOV is of 0.03 m, while the IFOV is 0.06 mm.}
Different sets of IR images are acquired and then processed as explained in the next coming sections.

\subsection{Thermal image processing}

Thermal imaging represents a visible version of thermal energy of the object in front of the camera in the form of an image. These images are then used for further inspection. During the last few years, several techniques based on infrared imaging have been proposed.  Dong et al. compared the registration methods and concluded that the registration method based on mutual information is more accurate but time-consuming, and hence are more suitable for thermal imaging sequence detection. However, methods based on automatic detection of features e.g. SIFT are more efficient, but have lower precision, and are suitable for the thermal imaging differential detection method \cite{Dong}. Similarly, Nandanwar et al. presented an optical inspection system for detecting bare PCB defects using multimodal approach \cite{Nandanwar}. Alaoui et al., used thermal signatures of components mounted on a PCB, in order to test them in production process  \cite{Alaoui}. More recently, Huang and Wei classified the defects and performed registration of same class \cite{Huang2019APD}. Also, Chaitali et al., presented the detection of faulty region of PCB using thermal image processing. They applied Principal Component Thermography (PCT) to compare both images \cite{Chaitali_detectionof}.

\begin{figure}[t]
  \centering
  \includegraphics[width=.75\textwidth]{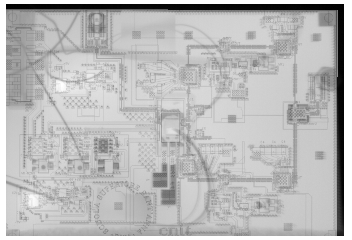}
    \caption{ Overlapping of layout image of the top layer of PCB design and IR images after registration.}
  \label{fig:overlap}
\end{figure}

This work emphasizes on identifying heat center taking into account heat distribution on the active in operation PCB by performing thermal imaging and later the spatial placement of these heat centers/traps were identified on layout files (GDSII/Gerber) of PCBs/ICs. The complete pipeline of processing is shown in Fig. \ref{fig:figure3}. After the detection of heat centers, the correspondence between IR and layout images (GDSII/Gerber) was established by performing geometrical transformation. Image registration is a mapping between two or more images both spatially (geometrically) and with respect to intensity. More precisely, registration is the correspondence or mapping that takes each spatial coordinate from the referral image and returns a coordinate for the defective image. Kim et al., presents a description and methods to perform registration \cite{kim}. Mathematically, registration is expressed as \cite{hines}:
\begin{equation}
I_{2} = gI_{1}(f(x_{1}; x_{2}));
\end{equation}
where $I_{1}$ and $I_{2}$ are two-dimensional IR and layout images (indexed by $x_{1}; x_{2}$),  $f : (x_{1}; x_{2}) \,\mapsto\, (x'_{1}; x'_{2})$ maps the indices of the distorted frame to match those of the reference frame, and $g$ is intensity or radiometric transform. At first, IR image and layout images (created using (GDSII/Gerber) files)  will be converted into gray image, followed by the extraction and matching of feature points of the two images, finally the transformation matrix is calculated to transform the test image into the same orientation and position as the layout image. In this paper, besides geometric transformation, control point selection tool is also used to set moving and fixed points between IR  and layout images. We have selected 4 corresponding pair point mainly belongs to heat centers as shown in Fig. \ref{fig:4point}. The control point pairs (the moving points in the IR and the fixed points in layout image) are defined by manually choosing the strongly correlated points as shown in Fig.  \ref{fig:4point}. The affine transformation is then applied for registering both layout and IR images. Fig. \ref{fig:overlap} shows the final overlapping of layout image of the top layer of PCB design and IR images after registration.\rev{ Image registration is performed in Matlab software tool. We implemented the image registration code using Matlab registration toolbox. In order to illustrate the steps involved in performing image registration using point mapping, the algorithm of the registration process is given in appendix}. After perfrecting image fusion, the thermal analysis of various electronic and photonic circuits is performed.

\section{Experimental results}
In order to provide the proof of concept, we have analyzed various optical and electronic circuits. Using FLIR A65 thermal imager, the IR profile (7.5 µm -14µm) was obtained. It is well-known that the thermogram of  conventional CMOS integrated circuits (IC) suffer from the effects of contrast due to high density of metals. The highly metallized part appears cold as compared to non-metallized area on the circuit. Therefore, in the absence of thermal energy, the IR profile may appear to have thermal contrast. Fortunately, the PIC requires few metal layers (<2) and thus conventional IR thermography can be applied to PIC without having need for lock-in. Therefore we have analyzed for a silicon based optical circuit with low metal density. The thermal profile reveals that using IR thermography, it is possible to distinguish heat sources on the chip with minimized effects of the contrast. We have also analyzed various electronic circuits (naked or bare and packaged die) for comparison with SOI \rev{(Silicon-on-Insulator)} devices. It was observed that CMOS circuits with high metal density provide low visibility of heat sources. Furthermore, we have also profiled populated PCBs and layout errors have been pointed using proposed IR-assisted fault monitoring technique presented earlier.  The thermograms of the following optical and electronic components/systems have been obtained and analyzed using the proposed analysis technique.

\begin{figure}[t]
 \centering
  \includegraphics[width=.75\textwidth]{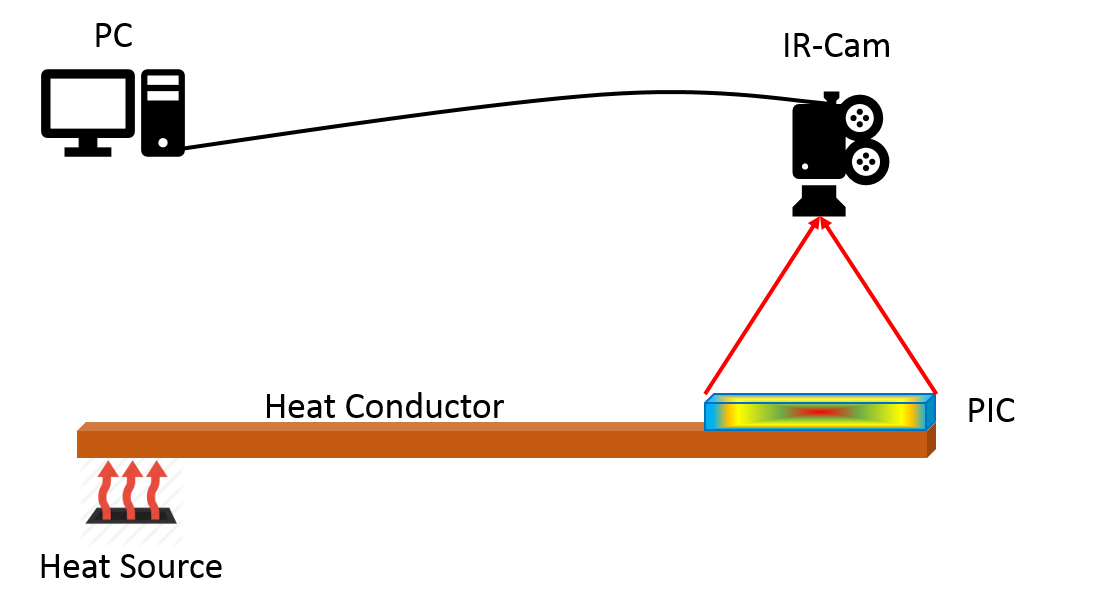}
    \caption{Experimental setup for thermography of SOI based photonic integrated circuit.}
  \label{fig:figure_4}
\end{figure}

\begin{figure}[t]
 \centering
  \includegraphics[width=.75\textwidth]{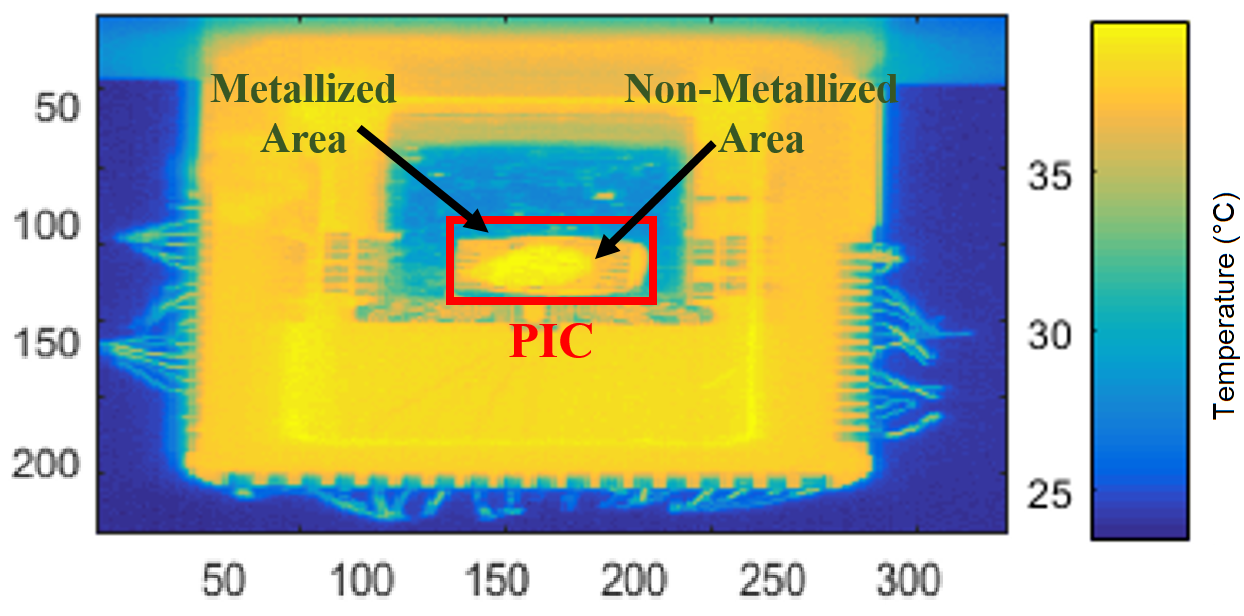}
    \caption{Thermal profile of SOI based PIC (size = 5mm x 5mm). The obtained IR image is 640 x 512 pixels but cropped for more focused visualisation.}
  \label{fig:figure_5}
\end{figure}

\begin{figure}[t]

\centering
 
  \includegraphics[width=.75\textwidth]{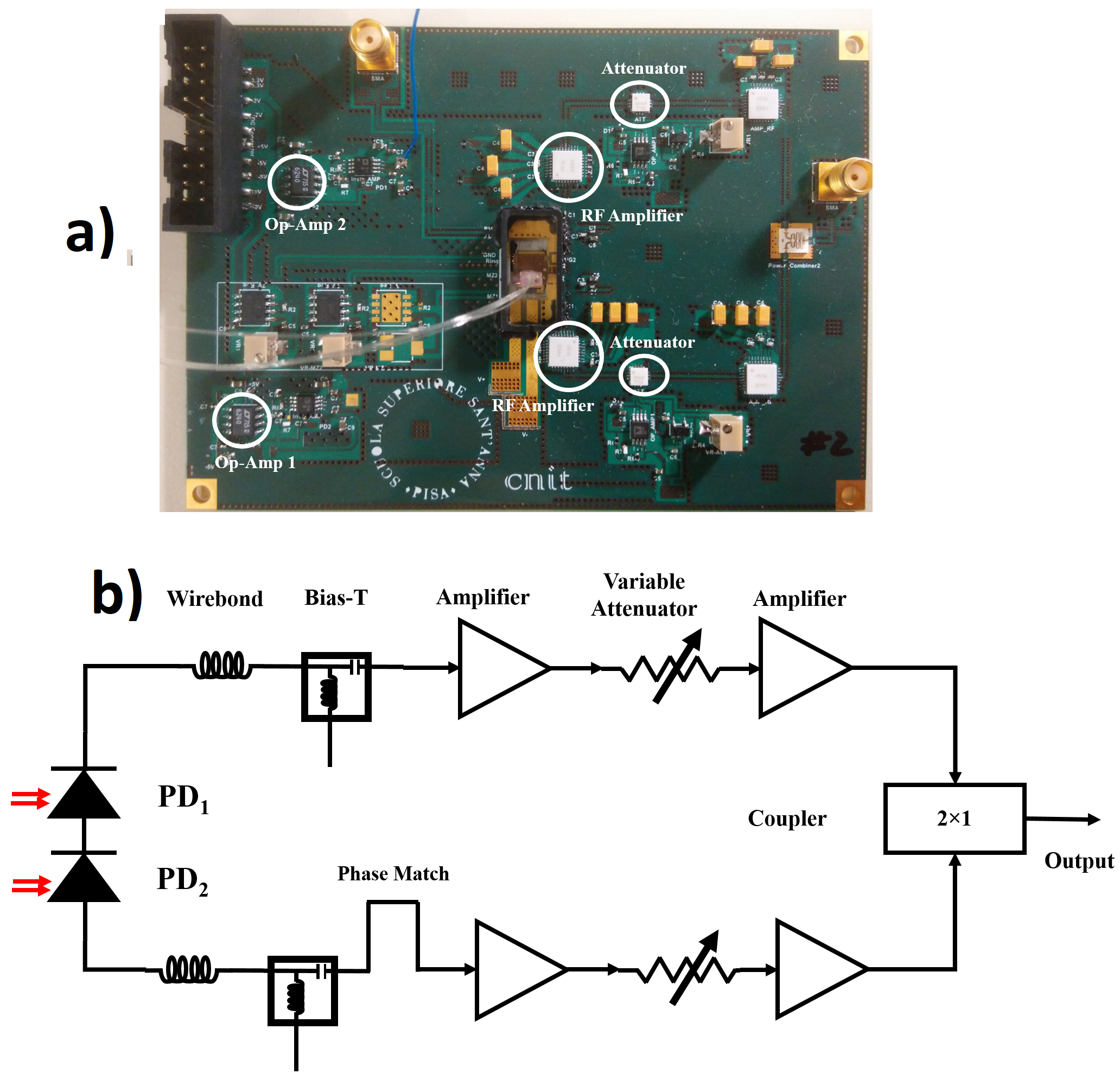}
    \caption{\rev{a) Photograph of photonics based RF \rev{beamforming} system (PCB size = 11.46cm x 8.22cm). b) Schematic of RF front-end for photonics based beamforming system. PD: Photodiode}}
  \label{fig:figure_6}
\end{figure}

\begin{figure}[t]
 \centering
  \includegraphics[width=.75\textwidth]{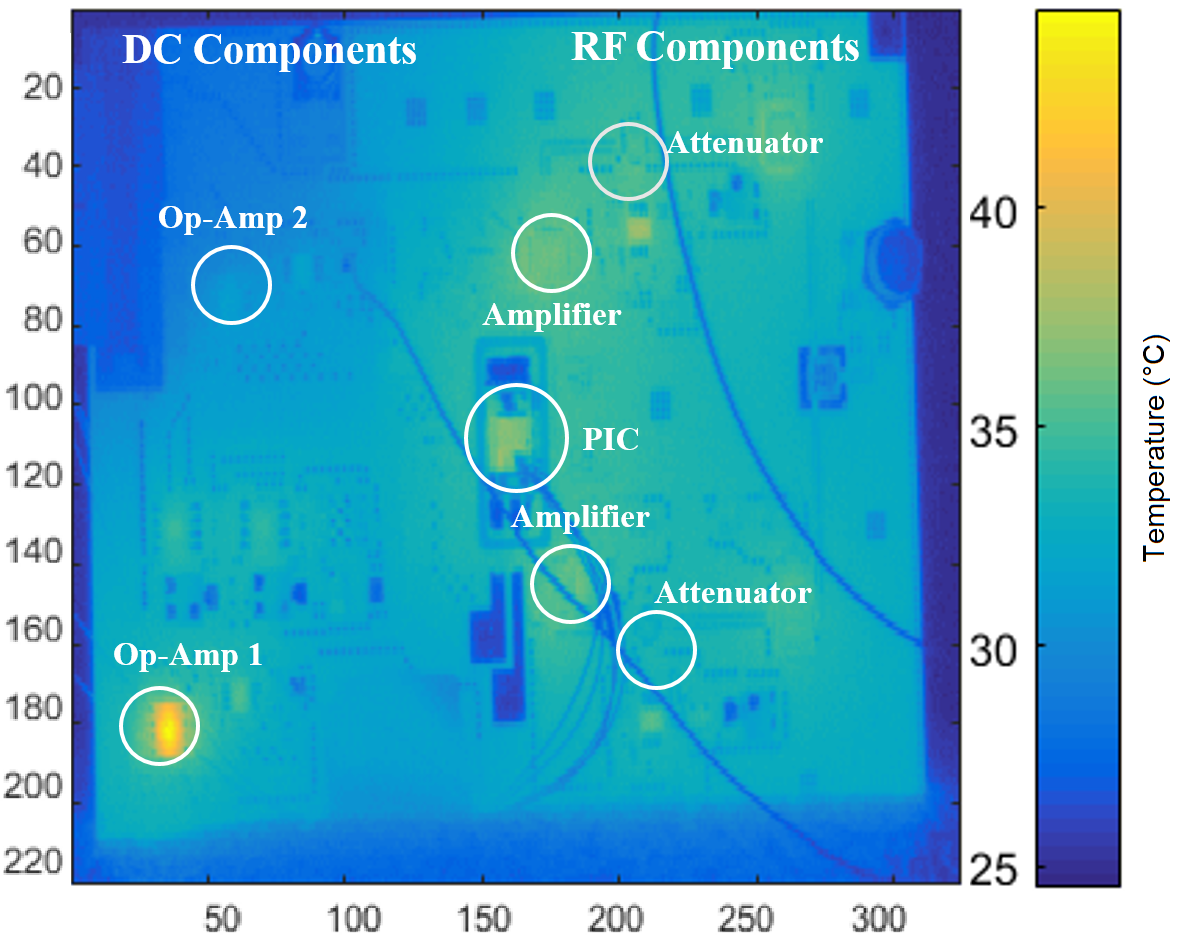}
    \caption{IR profile (top view) of photonics based RF beamforming system. The obtained IR image is 640 x 512 pixels but cropped for more focused visualisation.}
  \label{fig:figure_8}
\end{figure}

\subsection{Thermal analysis of SOI based optical circuits}
An SOI based optical circuit is used to acquire the thermal profile of optical circuits. The experimental setup of Fig. \ref{fig:figure_4} is used to characterize the photonic integrated circuit (PIC). A uniform heat source is constructed using a heating element and metallic heat conductor which can provide  a maximum temperature of 200°C. The source is thermally connected to the metallic conductor. Use of a metallic conductor guarantees uniform heating of the PIC, as the effective area of the heat source is smaller than that of the IC. Using FLIR A65, the thermal profile of the PIC is obtained.

The obtained gray scale thermal profile is processed using the image processing technique described in \cite{jalil2019preliminary}. From the thermal profiles presented in Fig. \ref{fig:figure_5}, it can be observed that the optical circuits contain lower metal density, hence weak heat sources can be observed. It provides the possibility of observing features comparable to the spatial resolution of the IR camera. From the thermogram of Fig. \ref{fig:figure_5}, it can also be observed that silicon is transparent to thermal radiation and the surface has very low absorption coefficient. The low absorption coefficient provides the possibility of uniform thermal emission from the buried silicon devices. Although the contrast issue still persists in the areas where metal density is high. Yet as the photonics circuits contain minimal metallization, the contrast in the thermal profile can be corrected by comparing it with the layout of the PIC. The presented thermal profiling setup can be used for monitoring passive/active optical devices on the chip. Combining thermal profiles with the layout (\rev{GDSII} format) of the IC, the setup can be used to identify thermally critical areas on the PIC . 

\begin{figure}[t]
  \centering
  \includegraphics[width=.75\textwidth]{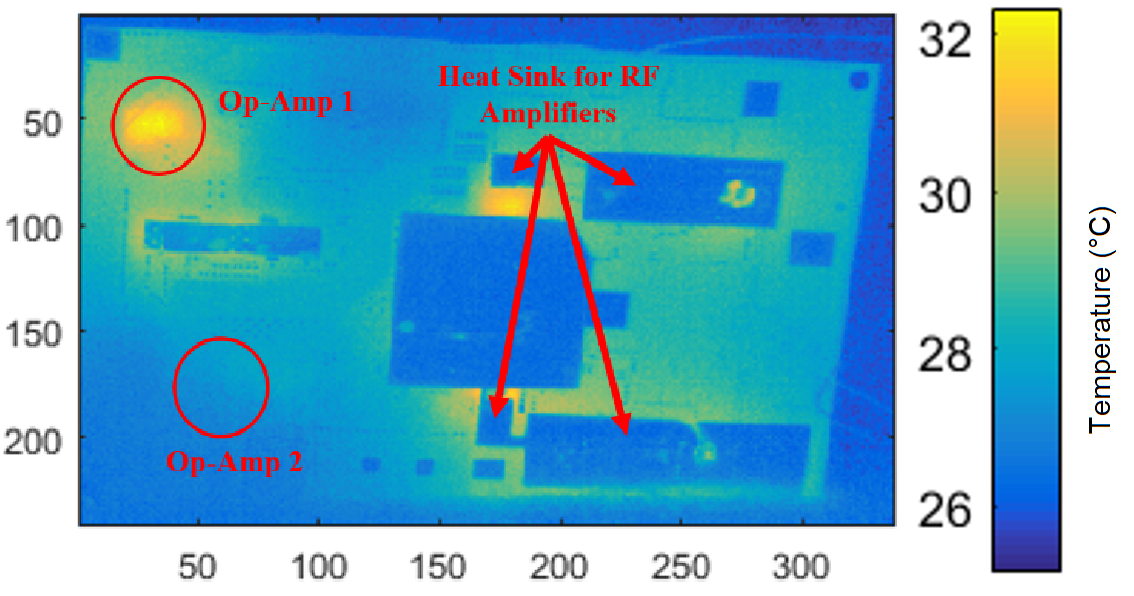}
    \caption{IR profile (bottom view) of photonics based RF beamforming system. The obtained IR image is 640 x 512 pixels but cropped for more focused visualisation.}
  \label{fig:figure_10}
\end{figure}

\subsection{Thermal analysis of a photonics based beamforming system}
As a further step in monitoring optical systems, a complete microwave photonics system is analyzed using the IR camera.  An RF beamforming system using silicon photonics based PIC is implemented on a 4-layer printed circuit board technology (PCB) as shown in Fig. \ref{fig:figure_6}(a). The PIC is housed in the center of the PCB over a thermo-electric cooler (TEC) glued to the PCB. The PCB contains DC and RF controls. In the right half plane, all of the RF  components are placed. The RF board contains four RF amplifiers along with two attenuators as shown in Fig. \ref{fig:figure_6}(b). Each RF component was provided a dedicated ground plane for conduction of heat. The bottom most layer is used for heat conduction. The heat pump is created using plated-through-holes (PTH) from top metal layer to bottom metal layer.  It is worth noting that all RF amplifiers are packaged and thus emissivity is dependent on the type of packaging. From the IR profile of Fig. \ref{fig:figure_8}, it can be observed that the heat is absorbed mostly in the solder mask layer. The solder mask  layer is not transparent to IR wavelengths and thus absorbs the incoming IR photons. This results in heating of the solder mask layer. The difference in the emissivity can be observed in the temperature contrast of IR profile. The solder mask and amplifier packaging has different emissivity as well as absorption coefficient, thus the heat profile varies around the packaged device. Using the suggested layout comparison technique, the difference in contrast can be used to determine the temperature boundaries for the RF components. Furthermore, the PTH via-holes are also visible. The PTH boundaries can be used to locate the heat centers in the respective metal layer. Although the surface absorption blurs the heat profile and thus makes it impossible to locate heat centers within the package. But as the feature size (comparable to wavelength) is not needed for PCB characterization, the provided resolution is sufficient for locating heat centers. 

In the left half plane of the PCB, all of the DC controls e.g. optical modulator bias, photo-diode (PD) bias and ring resonator bias are provided. Overall it seems that thermal image is blurred and heat is absorbed mostly in the solder mask layer (Fig. \ref{fig:figure_8}).

The IR profile shown in Fig. \ref{fig:figure_8}  reveals that there exists a heat center at the bottom left of PCB. Using comparison with the layout, it was noticed that the heat center represents an operational amplifier (LTC6240). The operational amplifier is used as a unity buffer to provide the bias current to the PD while monitoring it using an instrumentation amplifier (AD8421).  Fortunately, the PIC requires two monitoring circuits and both are placed on the same PCB with adequate isolation. The presence of only one heat center can be explained by comparing it with the layout of the PCB. It can be observed that the operational amplifier with missing heat center (Op amp 2 in Fig. \ref{fig:figure_6}) is conducting its heat using the underneath metal layers  while the absence of any metal underneath the Op-Amp1 results in the temperature increase. 

Such analysis proves that ability of the presented technique for HTOL (high temperature operating life) test where the IR profile of a circuit can be used to locate the thermal vulnerabilities of the circuit without subjecting it to a high operating temperature. Using the previously presented comparison technique, it is possible to precisely locate the layout errors which can be rectified by the designer at an early stage. Furthermore, by accurately mapping the IR profile, it is possible to predict the operating life of a device as a function of thermal stress. The critical areas where heat centers are located can potentially cause the break down of the device. Operating life can be calculated based on the location of heat center and its surrounding. 
The PIC is also visible in the IR profile of Fig. \ref{fig:figure_8}. Due to limited spatial resolution of IR camera, the distinctive temperature boundary is not visible. The bottom view of the PCB was also monitored for its IR profile. The IR profile (bottom-view of PCB) clearly presented the metallic contrast as the cold spots as shown in Fig. \ref{fig:figure_10}. It is interesting to observe that the heat centers are also visible from bottom view of the PCB. These heat center mostly correspond to areas where the value of DC current is on the order of 0.5A. The Op-amp1 appears also in the bottom view IR profile of the PCB. The limitation of this technique  is that the contrast due to high metal density limits the visibility of weak heat centers. The visibility of weak heat sources can be improved using lock-in thermography \cite{breitenstein2004thermal}.  The limitation of lock-in technique  is that the contrast due to high metal density limits the visibility of weak heat centers. 

\begin{figure}[b]
  \centering
  \includegraphics[width=.75\textwidth]{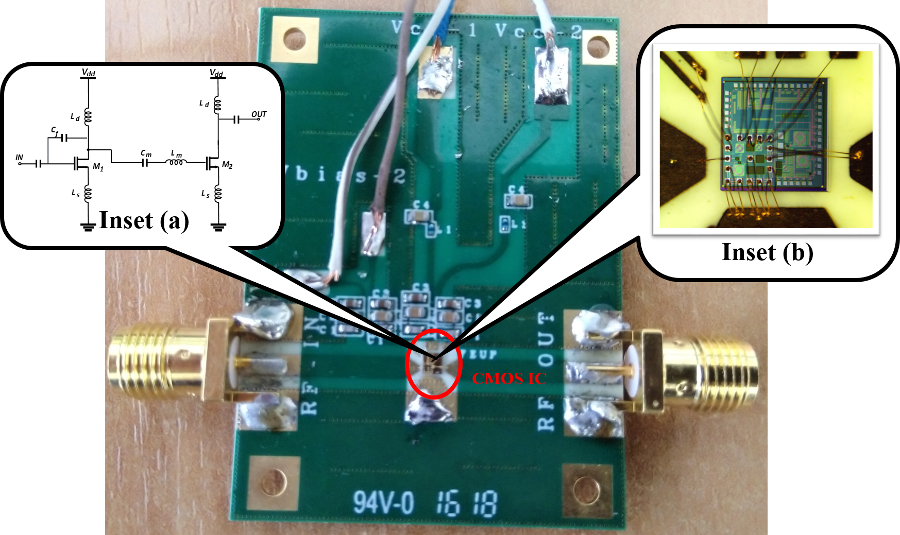}
    \caption{\rev{Photograph (visible range) of RF CMOS integrated circuit mounted on two-layered PCB (size = 2.5cm x 3cm) inset (a): Schematic of a cascade low noise amplifiers in CMOS, inset(b): Microscopic die photograph of CMOS IC in the visible range}}
  \label{fig:figure_11}
\end{figure}

\begin{figure}[t]
  \centering
  \includegraphics[width=.75\textwidth]{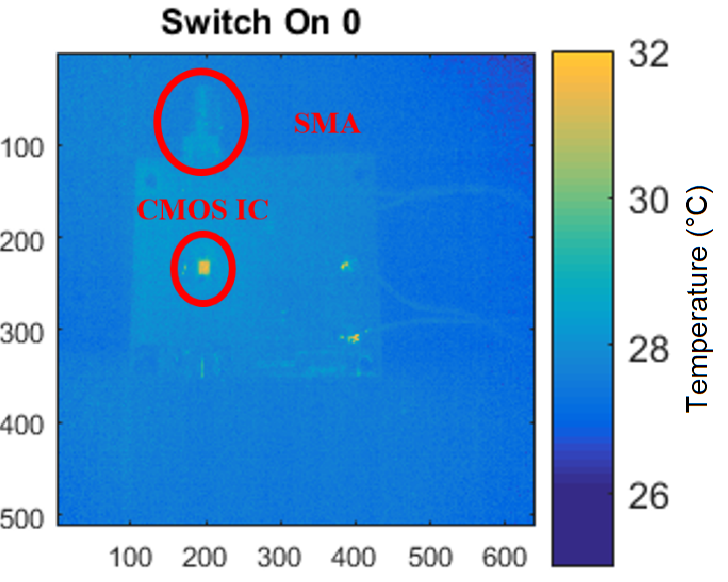}
    \caption{IR profile of RF CMOS IC mounted on a two-layered PCB. The obtained IR image is 640 x 512 pixels.}
  \label{fig:figure_12}
\end{figure}

\subsection{Thermal analysis of an RF CMOS IC}
For completeness of our investigation/effectiveness of our technique, we have also profiled a bare CMOS die mounted on a PCB. The visible and IR images of the PCB are presented in Fig. \ref{fig:figure_11} and Fig. \ref{fig:figure_12} respectively. The naked silicon die contains a low noise amplifier which dissipates a DC power of  36mW. The CMOS IC is mounted on two-layer Roger 4350b PCB. The thermograms are obtained for a duration of 240 seconds with alternating on/off cycles, in order to observe the heating/cooling of the CMOS IC within a specific time interval. 

The thermal profile of the CMOS IC reveals that the presence of high metal density smears the underlying heat centers and it becomes impossible to locate heat centers within the IC. Furthermore, the CMOS IC appears to be a single heating element where smaller features cannot be distinguished. From the microscopic image of CMOS IC (Fig. \ref{fig:figure_11}), it can be observed that there are several exposed metal surfaces on the top of the chip. Furthermore, the low noise amplifier (LNA) is placed on one side of the IC, thus the heat from LNA diffused through the SiO2 layers and over all image appears to be a blurred heat center. Also the size of CMOS IC is three times smaller than that of PIC, thus the spatial resolution of IR camera is also limited. The housed PCB appears to be at the minimum temperature. The invisibility of heat centers owes to the presence of larger exposed metal layer underneath for grounding purposes.  From the obtained heat profiles, it can be observed that proposed technique is not suitable for CMOS circuits with high metal density. Furthermore, an exposed ground plane underneath limits the visibility of heat sources across the circuit. In order to extend the presented technique to CMOS circuits, lock-in thermography is needed. Such technique will complicate the experimental setup and will be limited by the speed of IR camera. 

\subsection{Thermal analysis of a power amplifier}
\begin{figure}[t]
  \centering
  \includegraphics[width=.75\textwidth]{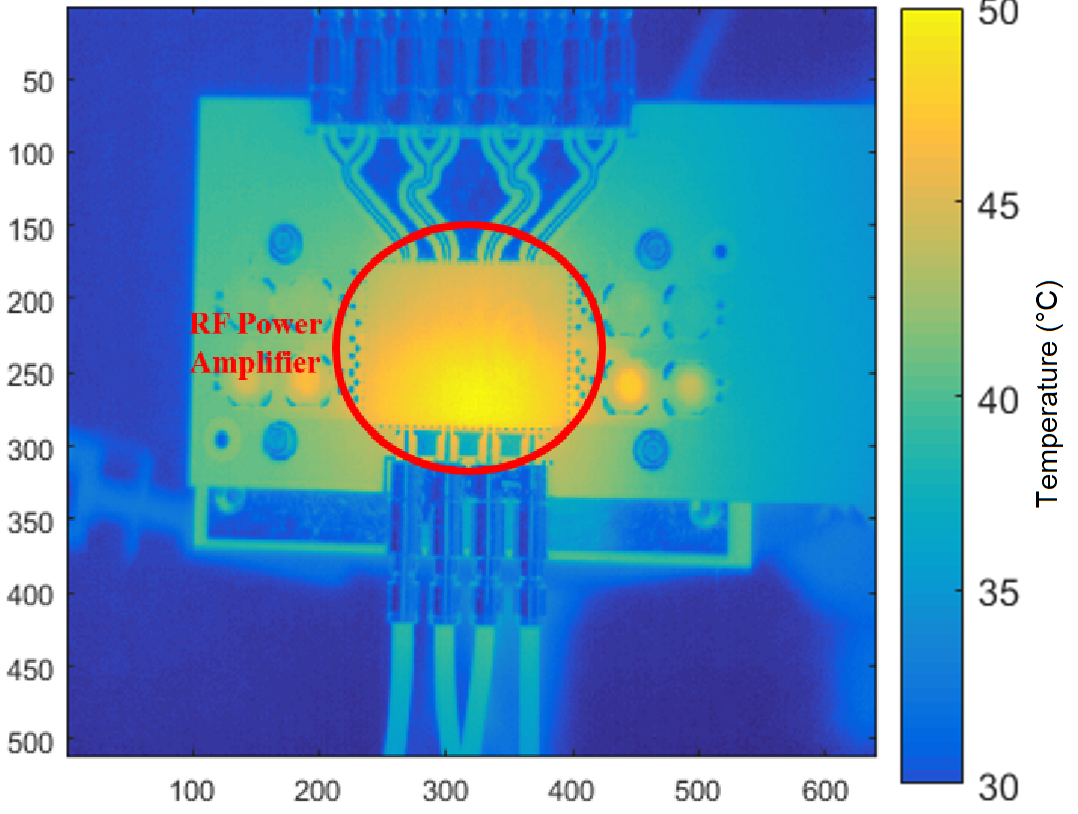}
    \caption{IR profile of an RF power amplifier. The obtained IR image is 640 x 512 pixels.}
  \label{fig:figure_13}
\end{figure}

In the case of CMOS IC, the low DC power dissipation resulted in the poor IR contrast. In order to further investigate the presented technique, a power amplifier was monitored using the same IR camera. The power amplifier operates over a bandwidth of 14GHz with a DC power consumption of 2W. The IR profile is presented in Fig. \ref{fig:figure_13}.
It can be noted that the themogram clearly presents the heat center within the host PCB. Furthermore, due to high RF power, the image contrast is improved and using the thermal profile, it is possible to determine the heat flow along the PCB. 
For CMOS based circuits, the IR thermography without a lock-in amplifier, is only suitable for high power applications. On the other hand, due to size and low metal density of PIC, such technique can be used for monitoring photonic circuits dissipating smaller DC power.

\subsection{Thermal analysis of digital circuits}
\begin{figure}[!b]
  \centering
  \includegraphics[width=.75\textwidth]{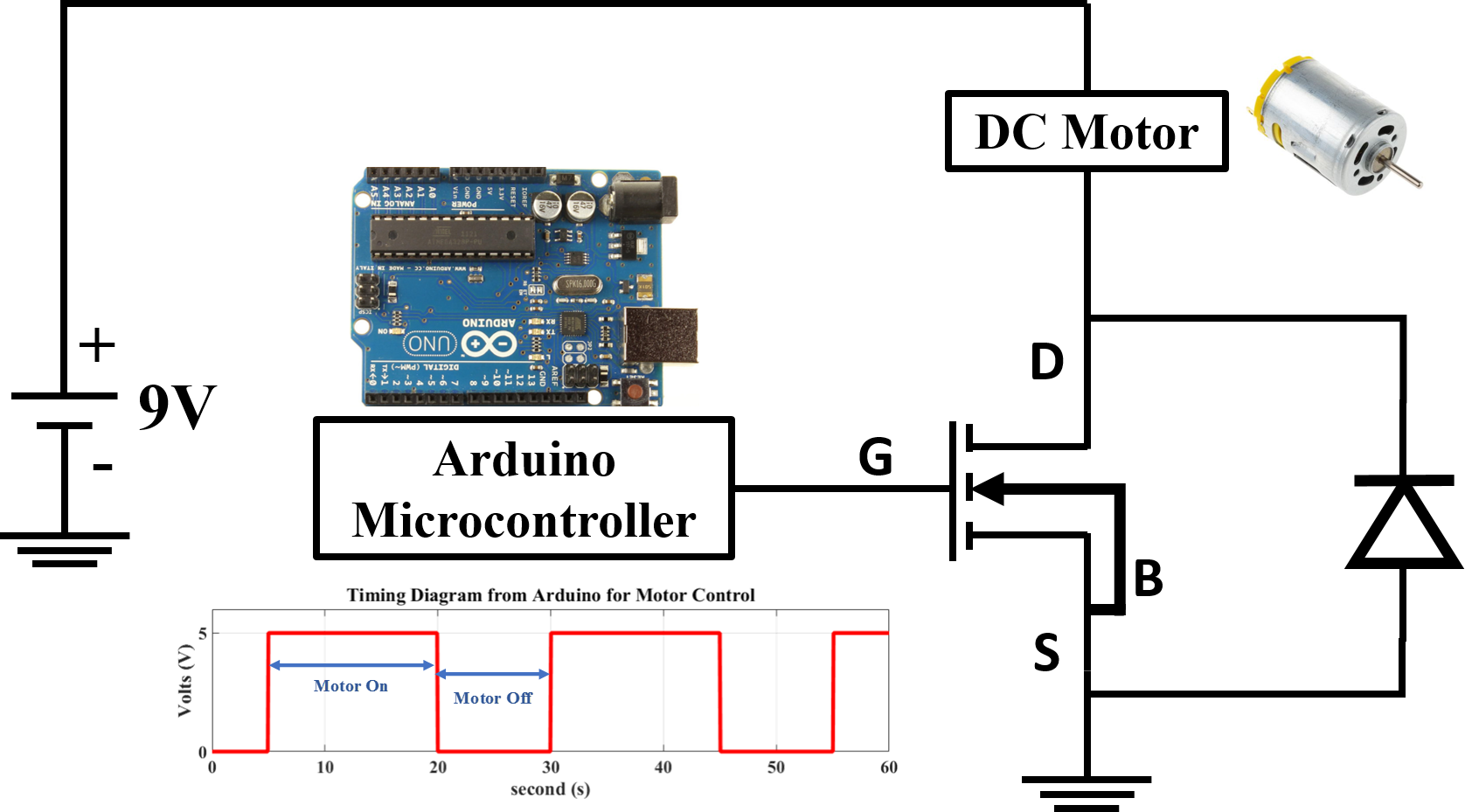}
    \caption{\rev{Schematic of a motor control implemented using Arduino Uno and high power MOSFET with terminals Drain (D),Source (S), Gate (G) and Bulk (B).}}
  \label{fig:figure_14}
\end{figure}

In order to complete the analysis, we have profiled digital circuits with active clocking operations. A commercial off the shelf Arduino Uno  was used to construct a motor controller \cite{arduino}. The schematic of the implemented circuit is shown in Fig. \ref{fig:figure_14}.  \rev{A DC motor is connected between the drain (D) and source (S) terminals of a high power MOSFET with a reverse biased protection diode.} The control signal from the micro-controller is programmed to operate in a pulsed format. The control signal (shown in Fig. \ref{fig:figure_14})\rev{ is applied to the gate (G) terminal of MOSFET which} allows the current flow to  the DC-motor  on the rising edge and the motor supply is cut-off at the trailing edge of the pulse.

\begin{figure}[t]
  \centering
  \includegraphics[width=.75\textwidth]{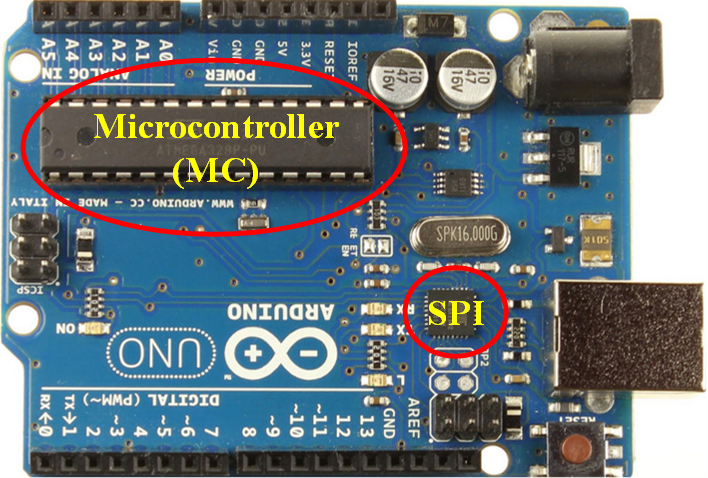}
    \caption{\rev{Photograph (visible range) of commercial Arduino board (size = 6.86cm $x$ 5.33cm). MC: Micro-controller, SPI: Serial-Parallel Interface}}
  \label{fig:figure_16}
\end{figure}

\begin{figure}[t]
  \centering
  \includegraphics[width=.75\textwidth]{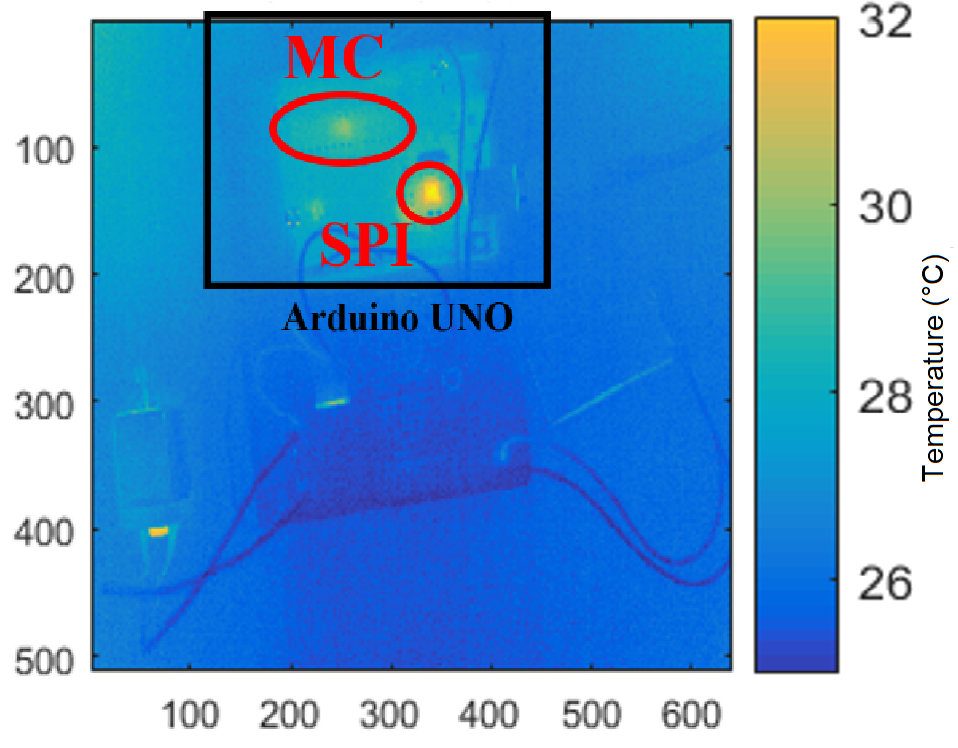}
    \caption{Thermal profile of motor control implemented using Arduino Uno.}
  \label{fig:figure_17}
\end{figure}

Arduino is programmed such that a continuous gating operation is performed during the thermal profiling. From the thermal profile presented in Fig. \ref{fig:figure_17}, it can be observed that Arduino Uno board contains one heat center. By using the layout comparison with the IR image, it was observed that Arduino board does not possess efficient heat conduction for ATmega328 serial-parallel interface (SPI) module  (Fig. \ref{fig:figure_16}). The microcontroller ATmega 168 operating at 20MHz can conduct heat efficiently. The setup was monitored for multiple on and off cycles. Furthermore, the MOSFET driving the motor at 9W shows minimal heating. The reason lies in the fact that the MOSFET is placed on a large heat sink and the package of MOSFET is not emitting any IR photons. This makes it harder to observe the true temperature of MOSFET.  From the IR profile, it can be observed that the presented technique is suitable for monitoring packaged digital circuits/system. The limitation of IR profiling comes from the fact that spatial resolution of the camera is limited, The analysis has provided the feasibility of the proposed technique for digital circuits.

\section{Conclusion}

This work has presented a novel layout comparison technique for successful IR profiling and locating fault centers within electronic and photonic integrated circuits / PCB technologies. For the first time, a comparison technique is developed such that it can compare the obtained IR profile with the generated layout of the device. The proposed technique can compare multiple layers provided that \rev {the heat centers are not superimposed or consequently only one layer is present at the heat center. The layers can only be identified  by comparing the location of heat centers with the corresponding layout. The areas which contain only one layer, it is possible to distinguish metal layers.} The proposed technique is experimentally verified using an existing microwave photonic system designed and fabricated for 5G communication. It was demonstrated that using the heat boundaries of components, it is possible to match the IR profile with the layout. This presents a novel tool for reliability testing of optical as well as electronic systems. Typically integrated as well as PCB based circuits are realized using the layout files in GDSII or Gerber format. These files are often accompanied with bill of material (BOM) and pick and place files. By using the knowledge of device/system layout, it is possible to map the fault vulnerabilities of a system and correctly locate the areas where design improvements are needed. The automation of this process can eliminate the requirement of an observer for identifying the fault lines within a circuit. The proposed technique can be effectively used to determine the vulnerable areas within a die or PCB during the HTOL test. The information on the thermal emission of a device helps in predicting the PoF and thereby creating a reliable model for prediction of operating life of a device. \rev{The precise location of heat centers within a circuit can help in determining the conventional failure mechanisms such as electromigration, time-dependent dielectric break down and hot carrier injection, Furthermore, this techniques provides the opportunity to monitor the optical circuits without subjecting them to very high thermal stress. Using lock-in thermography, a temperature resolution of 1mK is achievable. By incorporating the heat distribution profile with the predetermined failure mechanisms in silicon, it is possible to adapt conventional CMOS models for optical circuits which can predict the operating life of the device accurately. }
However it is important to mention that during this work, the proposed technique is only applied to PCB technologies due to limitation of resolution and focal length of the IR –CAM available in lab.  As a proof of concept, it is successfully presented that using the layout files, it is possible to map the IR images using the image processing techniques. The technique can be easily scaled to integrated circuit technology by just replacing the IR-CAM with a suitable camera with adequate features/specifications. The image processing technique is independent of the technology platform of the integrated circuits. 
Furthermore, the thermal profile of a PIC is presented. It is shown that due to lower metal density, it is possible to use conventional IR thermography for monitoring photonic systems. The recent advancements in the field of photonics will soon make the PIC a necessary component of electronic devices/systems, thus it will require an efficient way to predict the operating life of the device. Photonics being more susceptible to heat will require other forms of reliability tests. Using IR thermography, it is possible to monitor the electronic part of PIC without subjecting the device to increased thermal stress. 
This work is the first step towards fault modeling of photonic integrated circuits. \rev {The thermal profile of optical circuits is one of the key inputs to the operating life prediction models. Other variables such as humidity, radiation, material properties and fabrication tolerances are estimated using a detailed statistical analysis of silicon devices. Based on the aforementioned variables, the failure mechanisms are defined and a mathematical model is devised for predicting the operating life of a particular component/system.} Using the presented IR-assisted HTOL techiques, the next step will be analyzing the failure mechanisms within a PIC. As it was discussed that PIC are functionally different from conventional CMOS technologies, therefore it will be imperative to identify the fault mechanism within the IC. Active PIC such as lasers and photo-diodes are often implemented using III-V semiconductor materials. As a future work, it will be necessary to develop a comprehensive physics of failure such that it can encapsulate all photonic as well as electronic devices. The reliable IR profiling can provide added information for predicting the operating life a futuristic full functional microwave photonic system.

\section*{Appendix}
\textbf{Inputs = IR and layout images.}\\
\textbf{Output = Registered images.}\\
Load $I_{1}$ and $I_{2}$, two-dimensional IR and layout images.\\
Extract control points from both images $C1_{1,2,... x}$ and $C2_{1,2,... x} $.\\
\textbf{for} each point $ i = 1:x $ \\
    \qquad \textit{Find correspondence and match features.}\\
\textbf{end} \\
Apply Geometric Transformation $(T)$.\\
Output: Registered images.\\

\medskip
\section*{Disclosures}
The authors declare no conflicts of interest.

% Bibliography
%\bibliographystyle{acm}
%\bibliography{Bib_AITA.bib}

% Full bibliography added automatically for Optics Letters submissions; the following line will simply be ignored if submitting to other journals.
% Note that this extra page will not count against page length
%\bibliographyfullrefs{sample}

%Manual citation list
%\begin{thebibliography}{1}
%\bibitem{Zhang:14}
%Y.~Zhang, S.~Qiao, L.~Sun, Q.~W. Shi, W.~Huang, %L.~Li, and Z.~Yang,
 % \enquote{Photoinduced active terahertz metamaterials with nanostructured
  %vanadium dioxide film deposited by sol-gel method,} Opt. Express \textbf{22},
  %11070--11078 (2014).
%\end{thebibliography}

\end{document}